\documentclass[10pt, conference, letterpaper]{IEEEtran}
\usepackage{subfigure}
\usepackage{amsmath,amsfonts}
\usepackage{algorithmic}
\usepackage{algorithm}
\usepackage{array}
\usepackage[caption=false,font=normalsize,labelfont=sf,textfont=sf]{subfig}
\usepackage{textcomp}
\usepackage{stfloats}
\usepackage{url}
\usepackage{verbatim}
\usepackage{graphicx}
\usepackage{cite}
\usepackage{color}
\usepackage{booktabs}
\usepackage{threeparttable}
\usepackage{bbding}
\usepackage{pifont}
\usepackage{multicol}
\ifodd 1
\newcommand{\com}[1]{\textbf{\color{red}(COMMENT: #1)}} 
\else

\newcommand{\com}[1]{}
\fi

\def\fig{Fig.}

\def\eg{e.g.}
\def\ie{i.e.}

\def\name{TransformLoc }

\begin{document}

\newcommand{\mycustomsize}{\fontsize{23}{\baselineskip}\selectfont}

\title{
\mycustomsize{TransformLoc: Transforming MAVs into Mobile Localization Infrastructures in Heterogeneous Swarms}
}

\author{
    \IEEEauthorblockN{Haoyang Wang$^{1}$, Jingao Xu$^{2}$, Chenyu Zhao$^{1}$, Zihong Lu$^{3}$, Yuhan Cheng$^{1}$, Xuecheng Chen$^{1}$,}
    \IEEEauthorblockN{ Xiao-Ping Zhang$^{1}$, Yunhao Liu$^{2}$, Xinlei Chen$^{1,4,5 \dag}$}
    \IEEEauthorblockA{$^\dag$ Corresponding author}

    \IEEEauthorblockA{$^1$ Shenzhen International Graduate School, Tsinghua University, China; }
    \IEEEauthorblockA{$^2$ School of Software, Tsinghua University, China; $^3$ Harbin Institute of Technology, China; }
    \IEEEauthorblockA{$^4$ Pengcheng Laboratory, Shenzhen, China; $^5$ RISC-V International Open Source Laboratory, Shenzhen, China}

    \IEEEauthorblockA{Email: \{wanghaoyang0428, xujingao13, zhaocyhi, luzong2001, yhhncc, 1243566go\}@gmail.com, }
    \IEEEauthorblockA{xiaoping.zhang@sz.tsinghua.edu.cn, yunhao@tsinghua.edu.cn, chen.xinlei@sz.tsinghua.edu.cn}
}

\maketitle

\begin{abstract}
A heterogeneous micro aerial vehicles (MAV) swarm consists of resource-intensive but expensive advanced MAVs (AMAVs) and resource-limited but cost-effective basic MAVs (BMAVs), offering opportunities in diverse fields. Accurate and real-time localization is crucial for MAV swarms, but current practices lack a low-cost, high-precision, and real-time solution, especially for lightweight BMAVs. We find an opportunity to accomplish the task by transforming AMAVs into mobile localization infrastructures for BMAVs. However, turning this insight into a practical system is non-trivial due to challenges in location estimation with BMAVs' unknown and diverse localization errors and resource allocation of AMAVs given coupled influential factors. This study proposes TransformLoc, a new framework that transforms AMAVs into mobile localization infrastructures, specifically designed for low-cost and resource-constrained BMAVs. We first design an error-aware joint location estimation model to perform intermittent joint location estimation for BMAVs and then design a proximity-driven adaptive grouping-scheduling strategy to allocate resources of AMAVs dynamically. TransformLoc achieves a collaborative, adaptive, and cost-effective localization system suitable for large-scale heterogeneous MAV swarms. We implement TransformLoc on industrial drones and validate its performance. Results show that TransformLoc outperforms baselines including SOTA up to 68\% in localization performance, motivating up to 60\% navigation success rate improvement.
\end{abstract}

\section{Introduction}

Heterogeneous micro aerial vehicles (MAV) swarms offer transformative potential in conducting 4D (deep, dull, dangerous, dirty) tasks, especially quick response scenarios, such as search and rescue \cite{rashid2020socialdrone}, gas leak source detection \cite{mcguire2019minimal}, wildfire suppression\cite{khochare2021heuristic}, leveraging their inherent advantages of working scalability \cite{bertizzolo2020swarmcontrol}, flexibility \cite{wang2021lifesaving}, adaptability \cite{li2021physical}, etc. There are forecasts that the market size for MAV swarm-supported applications will reach \$ 279 billion by 2032 \cite{Global-Drones}.

\begin{figure}[t]
    \centering
        \includegraphics[width=1\columnwidth]{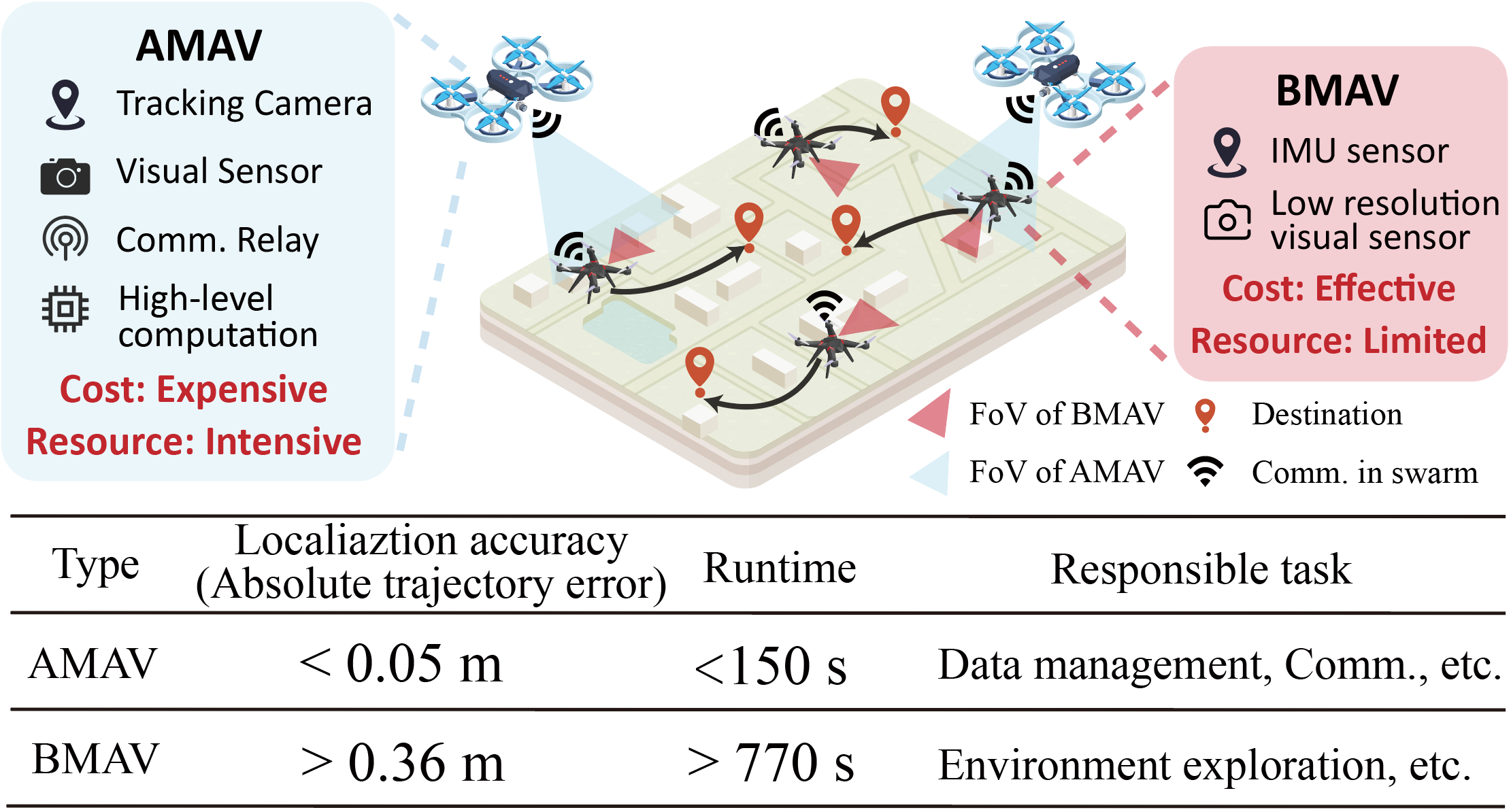}
    \vspace{-0.65cm}
     \caption{Introduction of heterogeneous MAV swarm. The AMAVs are resource-intensive and have a lower localization error and latency, while BMAVs are resource-limited, resulting in a high error and latency.}
    \label{intro}
    \vspace{-0.5cm}
\end{figure}

A heterogeneous MAV swarm typically consists of: 
$(i)$ a handful of advanced MAVs (AMAVs) equipped with intensive capabilities (eg. sensing, computing, etc.) yet expensive \cite{ruiz2016collaborative};
and $(ii)$ a larger group of resource-limited and cost-effective basic MAVs (BMAVs) \cite{chen2017design}, as shown in Fig.~\ref{intro} $\footnote{Results in Fig.1 and Fig.2a are measured with MH\_02 of EuRoC dataset\cite{burri2016euroc} and CCM-SLAM \cite{schmuck2019ccm}. AMAV is equipped with Intel(R) i7-8750H and 32G RAM, and BMAV is equipped with Cortex-A53 and 1G RAM.}$. 
The former primarily handles data management, communication, and complicated computation tasks, while the latter is dispersed into (hazard) target areas for data collection and environmental exploration \cite{li2022intelligent}. The AMAV-BMAV collaboration paradigm achieves an optimal balance between overall cap-abilities and cost, facilitating its widespread adoption \cite{trotta2018uavs}.

However, the diverse onboard resources also lead to unbalanced localization performance between AMAVs and BMAVs, which is the fundamental capability for flight control \cite{chi2022wi}, obstacle avoidance \cite{xu2020edge}, etc. Precisely, localization errors of the BMAV accumulate fast due to its noisy sensing data and low computational capability, preventing it from achieving accurate and real-time localization as the AMAV.

Unfortunately, current methods are not able to offer feasible solutions for BMAV localization in quick response scenarios, which can be divided into 2 categories:\\
\noindent $\bullet$ \textbf{Extra infrastructure based solutions}. These solutions utilize external devices to provide reference signals (\eg, GPS \cite{sharp2022authentication, xu2019ilocus}, RTK, radio \cite{li2021infocom_itoloc, chi2022wi}, or acoustic \cite{wang2022micnest}) for localization. They require 1) pre-deploying (densely) expensive localization infrastructure in the operation site and 2) line-of-sight connection between localization infrastructure and MAVs. However, in quick response scenarios like urban disaster relief, either requirement is hard to satisfy, which leads to localization failures \cite{chen2015drunkwalk, sun2022aim}.

\begin{figure}[t]
\setlength{\abovecaptionskip}{0.cm} 
\setlength{\belowcaptionskip}{-0.4cm} 
\setlength{\subfigcapskip}{-0.1cm}  
\centering
    \subfigure[Accuracy \& latency on BMAV]{
        \centering
            \includegraphics[width=0.8\columnwidth]{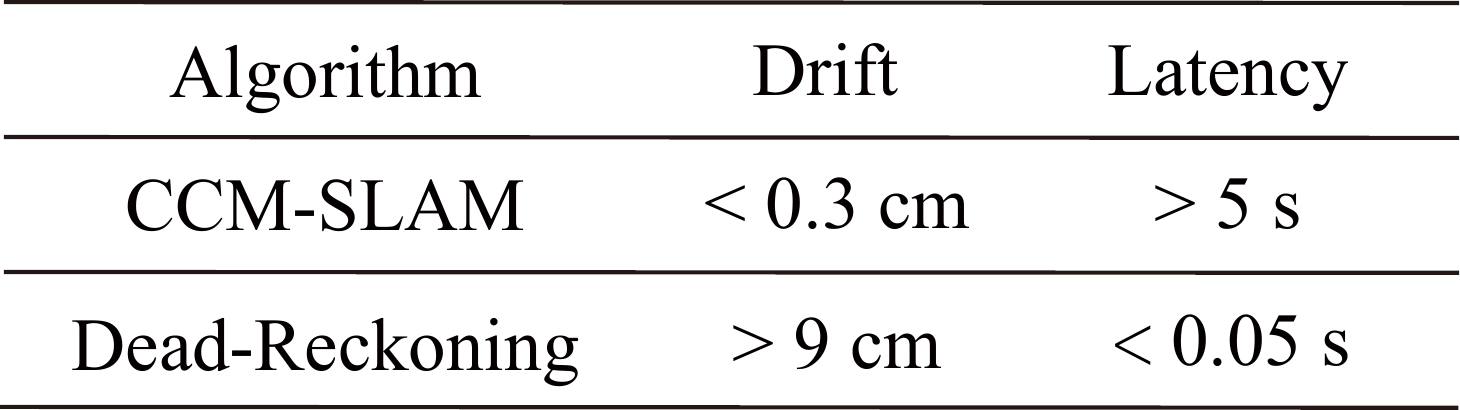}
    }
    \subfigure[Accuracy of AMAV observation]{
        \centering
            \includegraphics[width=0.8\columnwidth]{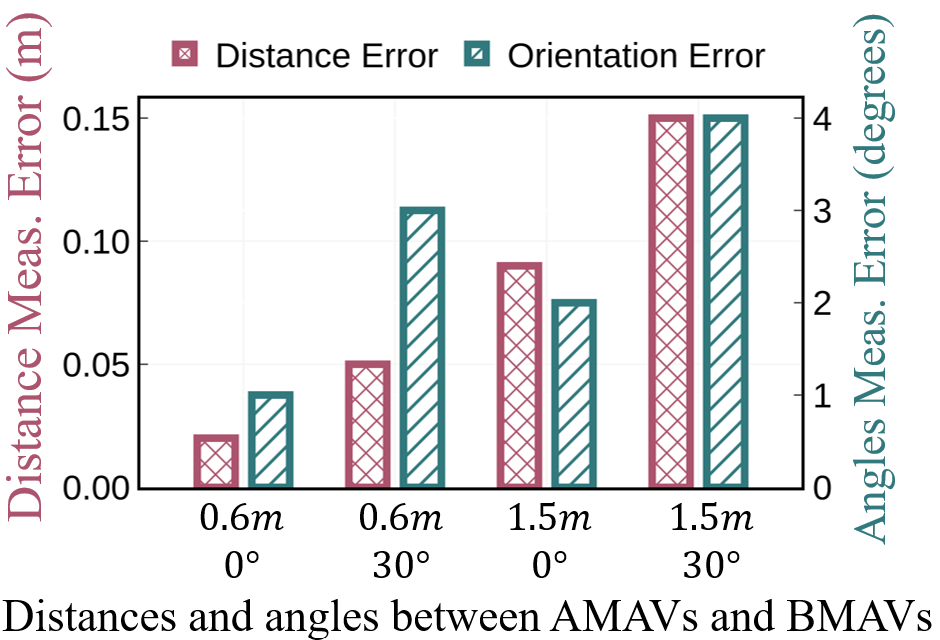}
    }
\caption{Motivating example. (a) BMAVs have limited on-board resources, resulting in a trade-off between accuracy and latency; (b) The accuracy of observation is affected by distances and angles between MAVs.}
\label{motivation}
\vspace{-0.6cm}
\end{figure}

\noindent $\bullet$ \textbf{Intra on-board sensor based solutions.}
These solutions leverage onboard sensors like cameras, IMU, LiDAR, and Radar, along with simultaneous localization and mapping (SLAM) techniques, to achieve high-precision location estimation\cite{lu2020milliego, dong2019infocom_pairnavi}. These methods require intensive on-board sensing and computing capabilities, which only work for AMAVs. In contrast, BMAVs' limited on-board resources result in significant accumulated errors or computation delays as illustrated in Fig.\ref{motivation}a \cite{xu2020edge}. 

Therefore, this paper aims to improve BMAVs' localization accuracy and efficiency given limited onboard sensing and computing capabilities without relying on any pre-deployed localization infrastructures. Our \textbf{key insight} is to transform a handful of AMAVs as \textit{mobile localization infrastructures} and offload their sensing and computing capabilities to support location estimation improvement of a larger group of BMAVs. Specifically, AMAVs provide external observations (with their visual sensors) for BMAVs to correct cumulative location estimation errors \cite{wang2016apriltag,xu2021mobisys_followupar}. 

However, turning this insight into a practical system is non-trivial, since two technical challenges have to be solved:

\noindent \textbf{$\bullet$ Unknown \& Diverse localization errors of BMAVs (C1).} On the one hand, accumulated location estimation errors may be diverse among BMAVs due to various reasons, such as various sensing noises, different operational conditions, etc. On the other hand, due to its limited field of view (FoV), each AMAV can only serve several BMAVs as localization infrastructure. 
Therefore, AMAVs should provide external observations for BMAVs with significant errors for estimation correction. 
However, online deriving BMAVs' localization errors is challenging due to the lack of static location references.
As a result, localization errors of BMAVs may keep accumulating without timely external observations from AMAVs.
Meanwhile, we need to consider how to correct BMAVs' location with AMAV's observations.

\noindent \textbf{$\bullet$ Resource allocation given coupled influential factors (C2).}
Assigning sensing and computing resources of a handful of AMAVs to a larger group of BMAVs as localization infrastructure involves optimization in high dimensional decision space. 
First, even if \textbf{C1} is solved, the presence of multiple AMAVs results in an exponential search space growth. 
Second, since both AMAVs and BMAVs are in constant motion, the observed BMAVs by each AMAV keep changing, necessitating dynamic adjustment of resource allocation strategies. 
Third, the localization effectiveness is affected by the accuracy of AMAVs' observations given various distances and bearing angles between AMAVs and BMAVs  (as shown in Fig. \ref{motivation}b), further complicating the resource allocation search space.
These factors influence resource allocation decisions in a coupled way, making it too complicated to run on MAVs.

To conquer these challenges, this paper presents TransformLoc, a novel collaborative and adaptive location estimation framework for heterogeneous MAV swarms. TransformLoc dynamically transforms AMAVs with intensive onboard capabilities into mobile localization infrastructures to support location estimation improvement for resource-constrained BMAVs. To achieve accurate and real-time localization for the entire swarm, TransformLoc only requires deploying intensive and expensive sensing and computing capabilities on a small number of AMAVs, without relying on external localization infrastructures.
Consequently, this framework facilitates the widespread deployment of swarm technology.

To address \textbf{C1}, we design an \textit{error-aware joint location estimation model}. This model designs an uncertainty-aided inference method to enable AMAVs to provide observations for BMAVs with larger errors at first. 
Subsequently, it integrates inaccurate estimation from BMAVs with discontinuous external observations from AMAVs to perform intermittent joint location estimation for BMAVs.

To address \textbf{C2}, a \textit{proximity-driven adaptive grouping-scheduling strategy} is proposed.
Initially, MAVs are dynamically grouped according to the principle of proximity, transforming the many-to-many resource allocation problem into multiple one-to-many resource allocation problems. 
Then, several steps lookahead about BMAVs are conducted to schedule each AMAV in a non-myopic manner, which also finds optimal distance and angle.

We evaluate the performance of TransformLoc and compare it with baselines including the state-of-the-art (SOTA) method through extensive experiments with both real-time testbed and physical feature-based simulations. 
Results show that TransformLoc is able to maintain BMAVs' average localization error under $1m$ with limited on-board resources in real-time, outperforming baselines up to $68\%$, which motivates up to $60\%$ navigation success rate improvement.

In summary, the main contributions are as follows:
\begin{itemize}
\item We propose TransformLoc, a new framework that dynamically transforms AMAVs into mobile localization infrastructures, enhancing localization accuracy and real-time performance for lightweight BMAVs.

\item We design an \textit{error-aware joint location estimation model} to boost the location estimation accuracy of BMAVs with discontinuous observation from AMAVs.

\item We design a \textit{proximity-driven adaptive grouping-scheduling strategy} to decouple the resource allocation issue given coupled influential factors. 

\item We validate our solution through in-field experiments on a real heterogeneous MAV swarm and large-scale physical feature-based simulations.
\end{itemize}

The remainder of the paper is organized as follows:
Section II provides an overview of TransformLoc, with detailed descriptions of the error-aware joint location estimation model in Section III and the proximity-driven adaptive grouping-scheduling strategy in Section IV. Section V showcases the implementation and evaluation. Sections VI and VII discuss the related work and influencing factors of the framework, respectively. Section VIII concludes TransformLoc. Appendix sections include detailed formula derivations of essential variables.

\section{Overview}\label{3}

\begin{figure*}[t]
    \centering
        \includegraphics[width=2.0\columnwidth]{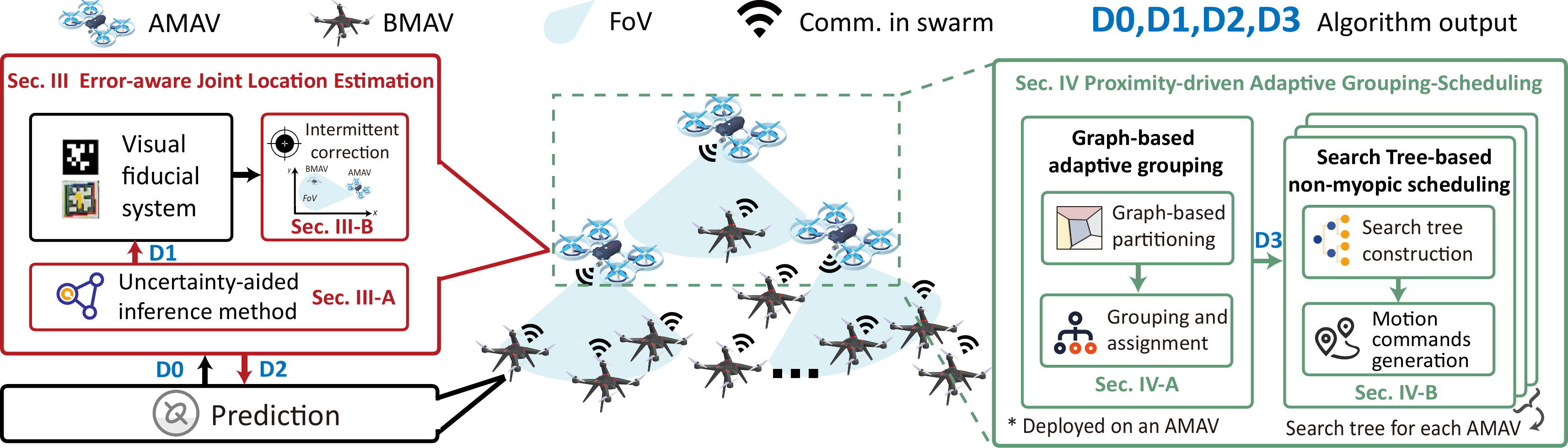}
        \vspace{-0.2cm}
    \caption{Illustration of \name framework. BMAVs estimate locations with noisy measurements. With the assistance of the uncertainty-aided inference method, AMAV generates discontinuous observations for BMAVs to perform intermittent joint location estimation. Subsequently, \name allocates resources of AMAVs by adaptive grouping and scheduling, which adaptively groups MAVs at first, and then schedules AMAVs in a non-myopic manner.}
    \label{architecture}
    \vspace{-0.6cm}
\end{figure*} 

\subsection{TransformLoc: Framework goal}
From the top perspective, we design and implement \name to transform AMAVs into mobile localization infrastructures for BMAVs. 
This adaptation is specifically crafted to cater to the economic and resource constraints of BMAVs, effectively addressing the inherent challenges presented by these limitations.
The goal of framework design is to answer the questions:

\noindent $\bullet$ How to optimize the location estimation of BMAVs with unknown and diverse localization errors? \name should infer the localization error of BMAV and utilize observations generated by AMAV to assist BMAVs, effectively reducing the localization error of BMAV.

\noindent $\bullet$ How to allocate sensing resources of AMAVs given coupled influential factors to assist BMAVs? \name should decouple the resource allocation problem and navigate the AMAVs in a non-myopic way, ensuring that the overall localization error of BMAVs remains low.

\subsection{Framework overview}
The architecture of \name is illustrated in Fig. \ref{architecture}. As seen, \name consists of two main components: 

\noindent $\bullet$ \textit{Error-aware joint location estimation model}.
Firstly, the BMAVs utilize the state model and motion commands to conduct the prediction of location at first, then transmit the yielded prior distribution of location to AMAVs (\textit{D0} in \fig \ref{architecture}).
Secondly, the uncertainty-aided inference method (\textit{Sec. III-A}) identifies when assistance is required for BMAVs (\textit{D1} in \fig \ref{architecture}). 
Finally, utilizing observations generated by visual fiducial system, AMAV intermittently corrects the location estimation of BMAVs (\textit{Sec. III-B}), then transmits the result posterior distribution of BMAVs' location and motion commands to BMAVs for following motions (\textit{D2} in \fig \ref{architecture}).

\noindent $\bullet$ \textit{Proximity-driven adaptive grouping-scheduling strategy}.
The number of AMAVs is limited, and the proximity-driven adaptive grouping-scheduling strategy is responsible to allocates resources of AMAVs to BMAVs. 
Firstly, this strategy groups MAVs adaptively utilizing a graph-based adaptive grouping method according to the relationship of AMAVs and BMAVs in the spatial dimension (\textit{Sec. IV-A}).
Secondly, utilizing grouping result (\textit{D3} in \fig \ref{architecture}), \name schedules each AMAV to allocate the sensing resource in a non-myopic manner by constructing a search tree for each AMAV.
Finally, \name generates motion commands for AMAVs to allocate sensing resources for BMAVs.

\section{Error-aware joint location estimation model}\label{4}

The BMAVs have unknown and diverse localization errors, complicating their location estimation with AMAV’s assistance.
In this part, we design an error-aware joint location estimation model based on Kalman filter and focus on localization error inference and joint location estimation of BMAVs. The main process is as follows:

\noindent $\bullet$ In order to enable AMAVs to provide observations for BMAVs with higher errors, we incorporate a measure of uncertainty to reflect the quality of BMAVs’ location estimation in Sec III-A.

\noindent $\bullet$ Meanwhile, we estimate the location of BMAVs with two coupled operations in Sec III-B: \textit{Prediction from state model} utilizes noisy motion measurements and the state model of BMAVs; \textit{Correction from observation} incorporates linearized observations generated by AMAVs.

\noindent The state model of BMAV and AMAV, observation model of AMAV are described in Appendix \ref{A1}. 

\begin{figure}[t]
    \centering    
        \includegraphics[width=1\columnwidth]{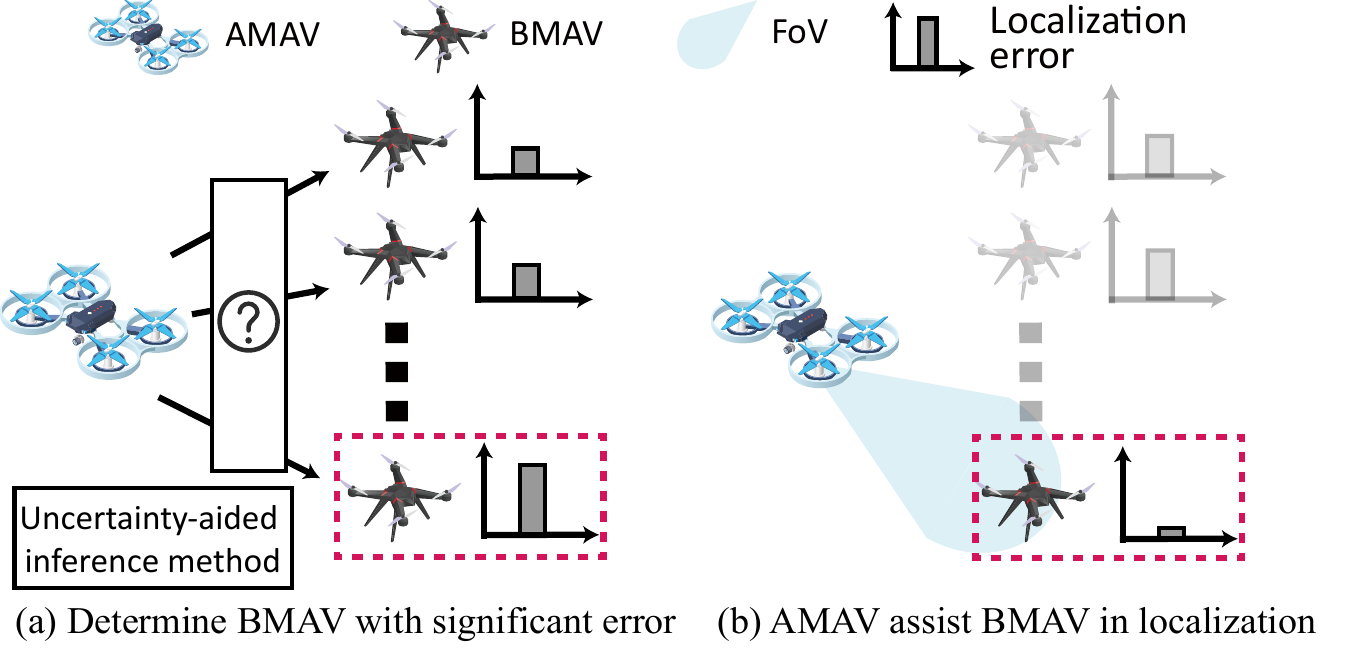}
        \vspace{-0.7cm}
    \caption{The error-aware joint location estimation model. (a) AMAV determines the BMAV with a higher error under the assistance of the uncertainty-aided inference method, and (b) generates observations for correction.}
    \label{error}
    \vspace{-0.55cm}
\end{figure} 

\subsection{Uncertainty-aided inference method}
The goal of \name is to efficiently allocate the sensing resources of AMAVs, thereby generating observations for BMAVs to mitigate localization errors.
The localization error of BMAV $B_i$ at time $t$ can be mathematically expressed as 
\begin{equation}
\psi(i, t) := E\left[{||y_{i, t}-\widehat{y}_{i, t}||}^2\right],
\label{le}
\end{equation}
where $y_{i, t}$ is actual location of $B_i$ and $\widehat{y}_{i, t}$ is location estimation of $B_i$.
By integrating the Kalman filter-based model for BMAVs' location estimation, we employ a metric of estimation uncertainty to gauge the accuracy of BMAVs' location estimations. This approach eliminates the necessity of having the actual locations of BMAVs to calculate the localization error.
In order to measure the BMAV's estimation uncertainty, we choose the trace of the covariance matrix of BMAV's location estimation\cite{ucinski2004optimal}, which can be mathematically expressed as  
\begin{equation}
\psi(i, t) = tr\left(\Sigma_{i, t}\right),
\label{ob}
\end{equation}
where $\Sigma_{i, t}$ is trace of the covariance matrix of $B_i$'s location estimation at time $t$.
This indicator measures the uncertainty of estimations, a lower value indicates greater certainty.
For BMAVs exhibiting varying degrees of localization errors (Fig. \ref{error}a), the AMAV generates observations tailored to assist BMAVs with more substantial errors (Fig. \ref{error}b).

\subsection{Joint estimation of BMAVs' location}
The joint location estimation framework of BMAVs includes two operations: \ding{172} prediction from the state model and noisy velocity measurement of BMAVs, which calculates the prior distribution of BMAVs' location; \ding{173} correction from the observation of AMAVs, which calculates the posterior distribution of BMAVs' location. The detail of joint estimation of BMAV's location is illustrated in Algorithm \ref{estimation}.

\textbf{Prediction of BMAVs from state model}.
This process is outlined in lines 1-3 of Algorithm \ref{estimation}. 
It leverages the motion model of BMAVs, incorporating the noisy motion measurement $\dot{v}_{i, t-1}$ and the location estimation $\boldsymbol{\hat{y}_{i, t-1}}$ of BMAV $B_i$ at time $t-1$ to compute the prior distribution and covariance matrix of $B_i$'s location estimation, denoted as $\boldsymbol{{y_{i, t}}^-}$ and ${\Sigma_{i, t}}^-$.

\textbf{Correction of BMAVs from observations}. 
This process is delineated in lines 4-11 of Algorithm \ref{estimation}.
When BMAV $B_i$ is within the Field of View (FoV) of AMAV $A_j$ at time $t$, this process employs the observation $z_{i, j, t}$ and the prior distribution of $B_i$'s location estimation to compute the posterior distribution and covariance matrix at time $t$, denoted as $\boldsymbol{{y_{i, t}}^{+}}$ and ${\Sigma_{i, t}}^+$. The variable $\eta$ in line 8 represents a normalization constant. 
If $A_j$ generates an observation for $B_i$, the location estimation $\boldsymbol{\widehat{y}_{i, t}}$ and covariance matrix of estimation $\Sigma_{i, t}$ for $B_i$ at time $t$ are drawn from the posterior distribution $\boldsymbol{{y_{i, t}}^+}$ and ${\Sigma_{i, t}}^+$; otherwise, they are drawn from the prior distribution $\boldsymbol{{y_{i, t}}^-}$ and ${\Sigma_{i, t}}^-$.

\begin{algorithm}[t]
	\renewcommand{\algorithmicrequire}{\textbf{Input:}}
	\renewcommand{\algorithmicensure}{\textbf{Output:}}
	\caption{$A_j$ assists $B_i$ for localization using the noisy motion measurement and the observation.} 
    	\begin{algorithmic}[1]
            \REQUIRE Location estimation of $B_i$ at time $t-1$, $\boldsymbol{\widehat{y}_{i, t-1}}$; covariance matrix of estimation at time $t-1$, $\Sigma_{i, t-1}$; noisy motion measurement, $\dot{v}_{i, t-1}$; location of $A_j$, $\boldsymbol{x_{j, t-1}}$; motion command of $A_j$, $u_{j, t-1}$.
            \ENSURE Location estimation of $B_i$ at time $t$, $\boldsymbol{\widehat{y}_{i, t}}$; covariance matrix of estimation at time $t$, $\Sigma_{i, t}$.\\
                \textit{\textbf{\% Prediction of BMAV from state model}}
    		\STATE Update prior distribution of $B_i$'s location at time $t$,\\ $\boldsymbol{y_{i, t}}^{-} =\int p\left(\boldsymbol{y_{i,t}} \mid \boldsymbol{y_{i,t-1}}, \dot{v}_{i, t-1} \right) \boldsymbol{\hat{y}_{i,t-1}} d \boldsymbol{y_{i,t-1}}$;
          	\STATE Update covariance of estimation ${\Sigma_{i, t-1}}^{-}$ from $\boldsymbol{y_{i, t}}^{-}$;
                \STATE Update $\boldsymbol{\widehat{y}_{i, t}}$ and $\Sigma_{i, t}$, from $\boldsymbol{y_{i,t}}^{-}$ and ${\Sigma_{i, t}}^{-}$;\\
                  \textit{\textbf{\% Correction of BMAV from observations}}
                \STATE Update the FoV of $A_j$, $F_{j, t}$ according to Eq.(\ref{F_t});
    		\IF{$B_i$ in the FoV of $A_j$}
                \STATE Update observation $\boldsymbol{z_{i,j,t}}$ according to Eq.(\ref{z});
                \STATE Linearize the observation according to Eq.(\ref{linearize});
    		\STATE Update posterior distribution of $B_i$' location at $t$, \\$\boldsymbol{y_{i, t}}^{+}=\eta p\left(\boldsymbol{z_{i,j,t}} \mid \boldsymbol{y_{i,t}}\right) \boldsymbol{y_{i, t}}^{-}$;
                \STATE Update covariance of estimation ${\Sigma_{i, t}}^{+}$ from $\boldsymbol{y_{i, t}}^{+}$;
    		\STATE Update $\boldsymbol{\widehat{y}_{i, t}}$ and $\Sigma_{i, t}$ from $\boldsymbol{y_{i,t}}^{+}$ and  ${\Sigma_{i, t}}^{+}$;
    		\ENDIF
	\end{algorithmic} 
\label{estimation}
\end{algorithm}

\section{Proximity-driven adaptive grouping-scheduling strategy}\label{4}
The resource allocation of AMAVs is influenced by coupled factors, involving optimization in a high-dimensional decision space.
In this part, we design a \textit{proximity-driven adaptive grouping-scheduling} strategy to allocate resources of AMAVs to assist BMAVs in localization. The main process is as follows:

\noindent $\bullet$ This strategy first dynamically groups the AMAVs and BMAVs according to the proximity in spatial domain based on the Voronoi diagram. This step transforms the many-to-many resource allocation problem into multiple one-to-many resource allocation problems.

\noindent $\bullet$ Following that, to strategically plan each AMAV in a non-myopic manner and determine the optimal observational distance and angle, this strategy constructs a search tree for each AMAV, incorporating several steps of lookahead regarding BMAVs.

\noindent The mathematical formulation of the resource allocation problem is described in Appendix \ref{A2}.

\subsection{Graph-based adaptive grouping}
The BMAVs are located in various locations with varying localization errors. 
When generating observations for BMAVs, a single AMAV is limited by its location and waste sensing resources by moving between different BMAVs. 
In this section, we dynamically group BMAVs so that each AMAV can focus its sensing resources on one group. 
Grouping BMAVs and assigning them to different AMAVs pose a combinatorial optimization challenge that becomes inherently difficult to solve with a substantial number of BMAVs and AMAVs, owing to its NP-hard nature.
To tackle this challenge, \name employs two key operations:

\noindent $\bullet$ \name initially divides the entire area into non-overlapping regions according to the locations of AMAVs. 

\noindent $\bullet$ Subsequently, each AMAV allocates sensing resources for BMAVs within the nearest region for a duration of $\delta$, which is the control command interval for BMAVs. 

\noindent As a result, all of BMAVs are categorized into non-overlapping groups, with distinct groups being assigned to different AMAVs. The details are provided below.

\begin{figure*}[t]
    \centering
        \includegraphics[width=2\columnwidth]{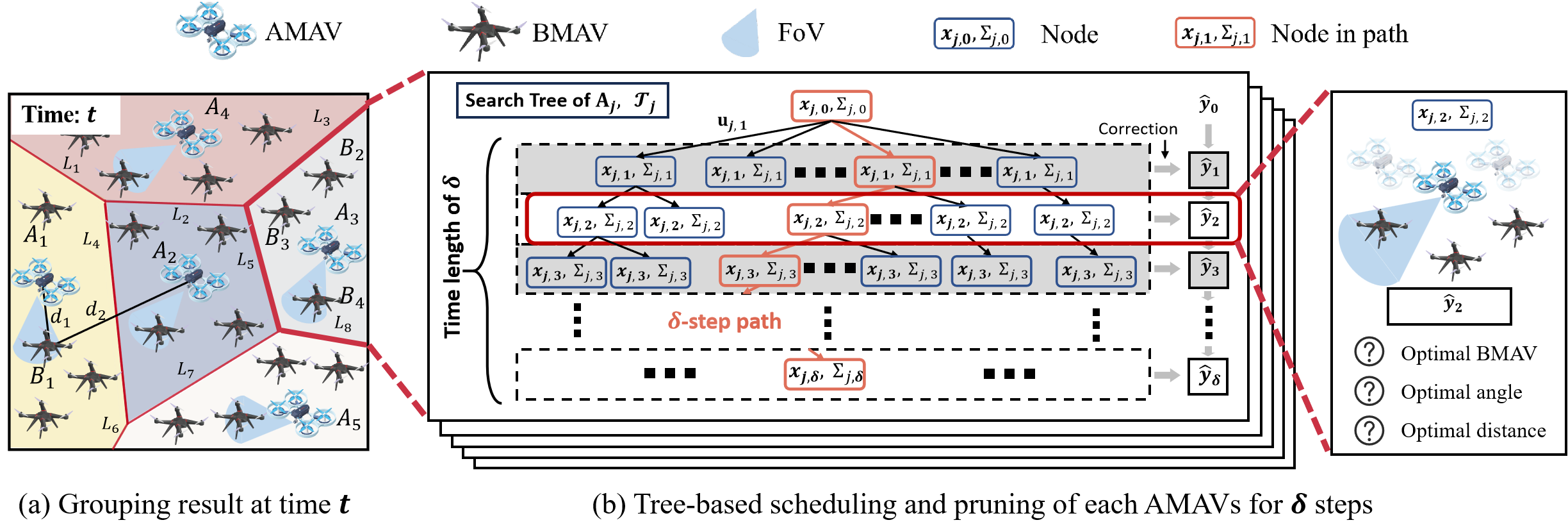}
    \vspace{-0.35cm}
    \caption{The proximity-driven adaptive grouping-scheduling strategy. This strategy groups MAVs based on the Voronoi diagram to decouple the resource allocation problem. 
    Then, it constructs search trees for each AMAV by involving several steps lookahead about BMAVs, resulting in $\delta$-step paths for AMAVs. 
    AMAV is scheduled in a non-myopic way to assist BMAVs with significant errors in an optimal distance and angle. 
    $x_{j, 0}$ is state of $A_j$, $\Sigma_{j, 0}$ is covariance of assigned BMAVs' estimation maintained by $A_j$, and $\hat{y}_0$ is BMAVs' estimation location.}
    \label{grouping-scheduling}
    \vspace{-0.4cm}
\end{figure*} 

\textbf{Graph-based region partitioning.} 
The partitioning approach is based on the Voronoi diagram \cite{zhang2021emp}. 
Under this scheme, each AMAV is associated with a region encompassing points whose distance to the given AMAV is less than or equal to their distance to any other AMAV. 
Fig.\ref{grouping-scheduling}a presents a depiction of this partitioning method with five AMAVs and several BMAVs. 
The result is represented by region boundaries denoted as \textit{L1} - \textit{L8}, which are separated by perpendicular bisectors of neighboring AMAVs. 
Any BMAV located within the region of $A_1$ is closer to it than to any other AMAV (\ie, \textit{d1} \textless \textit{d2} in Fig.\ref{grouping-scheduling}a).

\textbf{Grouping and assignment of MAVs.}
BMAVs within the designated region of an AMAV are organized into groups, and the AMAV allocates its sensing resources exclusively to assist in their localization for a duration of $\delta$. 
The group renews after an interval of $\delta$. 
Importantly, the boundaries of an AMAV's region are solely determined by the locations of its neighboring AMAVs and can be computed by identifying the perpendicular bisectors between adjacent AMAVs. 
If no BMAVs are present within an AMAV's region, it allocates sensing resources to all BMAVs over the duration of $\delta$.

\subsection{Search tree-based non-myopic scheduling}\label{4.4}
In this section, we present a methodology for integrating a search tree-based scheduling strategy into the non-myopic resource allocation of AMAVs.
The primary procedure unfolds as follows:

\noindent $\bullet$ BMAVs receive commands at discrete time intervals of $\delta$, enabling the acquisition of BMAVs' motion commands within each $\delta$ interval for AMAVs. 

\noindent $\bullet$ These motion commands are subsequently employed to plan trajectories for AMAVs, incorporating a $\delta$-step lookahead in coordination with the movements of BMAVs.

\textbf{Search tree construction.}
We construct search trees for each AMAV. 
As Fig. \ref{grouping-scheduling}b shows, we construct a search tree $\mathcal{T}_j$ for $A_j$ as an example. 
$\mathcal{T}_j$ includes a set of candidate trajectories $A_j$ can take, starting from an initial location and covariance pair $(\boldsymbol{x_{j,0}}, \Sigma_{j,0})$, where $\boldsymbol{x_{j,0}}$ is the starting location of $A_j$, and $\Sigma_{j,0}$ is the initial covariance of the location estimations of the BMAVs assigned to $A_j$.
The nodes of the search tree at level $t \leq \delta$ correspond to reachable locations for $A_j$ and are denoted as $(\boldsymbol{x_{j,t}}, \Sigma_{j,t})$. 
The AMAV measures the distance and angle for observable BMAVs at each location to determine an optimal distance and angle to generate observations for BMAVs with significant error.
We discretized the control space of the AMAV, $A_j$ has a finite set of control options $\mathcal{U}$, with an edge for each option $\boldsymbol{u_{j, t}}$ starting at node $(\boldsymbol{x_{j,t}}, \Sigma_{j,t})$ and leading to node $(\boldsymbol{x_{j,t+1}}, \Sigma_{j,t+1})$ by evaluating state model of AMAV and Algorithm \ref{estimation}.
Then estimation location of corresponding BMAVs $\boldsymbol{\hat{y}_t}$ is computed by evaluating motion model of BMAV, observation model of AMAV, and Algorithm \ref{estimation}.

\textbf{Motion commands for AMAV.}
Upon finishing the construction of the search tree for an AMAV $A_j$, we choose the node at level $\delta$ with the minimum value of $tr(\Sigma_{j, \delta})$.
Subsequently, through backtracking on this node, we plan the trajectory of $A_j$ navigate it for a duration of $\delta$.

\subsection{Scheduling of BMAV} 
\name utilizes location estimation to navigate BMAVs to their destinations. 
To achieve this, we design a lightweight planning algorithm based on an artificial potential field, ensuring BMAVs can avoid collision. The key points are as follows: 

\noindent $\bullet$ \name maintains BMAVs' location estimation and generates motion commands with an interval of $\delta$ based on the distance to their destination. 

\noindent $\bullet$ BMAVs move within a field of forces, the destination attracting them through a force proportional to the distance. The Wall and other BMAVs generate repulsive forces which repel the BMAV.
This approach enables BMAVs to dynamically adjust motion when nearing the destination, decreasing velocity and increasing navigation success.
\section{Evaluation}\label{5}
\subsection{Implementation and Methodology}

\begin{figure}[t]
    \centering
        \setlength{\abovecaptionskip}{0.cm}
        \includegraphics[width=1\columnwidth]{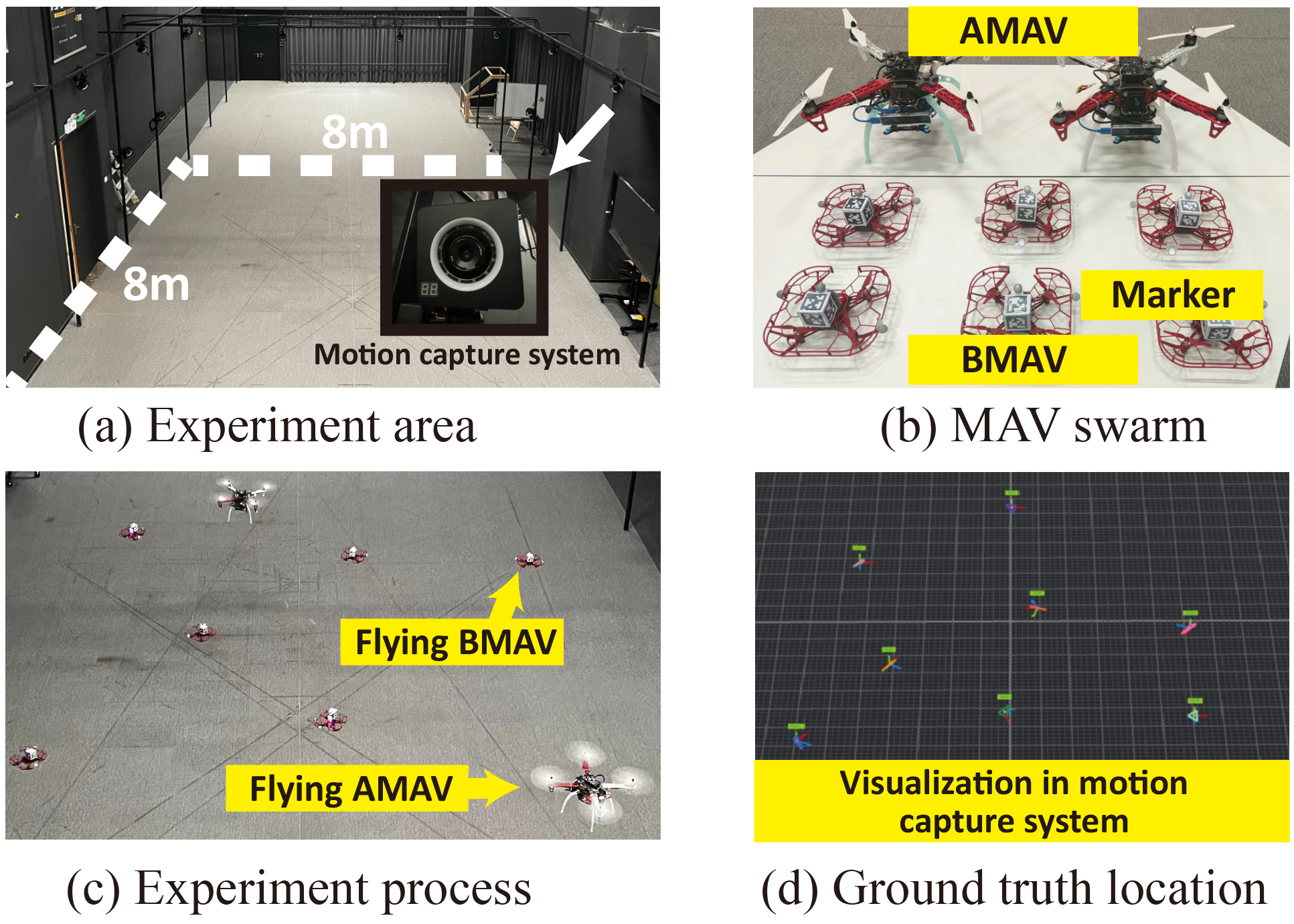}
    \caption{Experiment area and heterogeneous MAV swarm implementation. The AMAVs utilize a tracking camera to localize themselves and visual sensors to generate observation. BMAVs mount AprilTag markers for recognition and localization. The ground truth is obtained from the motion capture system. The experiment area has a size of $8m \times 8m$.}
    \label{setup}
    \vspace{-0.6cm}
\end{figure} 

\textbf{Testbed Implementation.}
As Fig. \ref{setup} illustrated, we implemented the TransformLoc based on DJI Robomaster TTs (BMAVs) and industry drones (AMAVs) built on Pixhawk which is one of the most widely used autopilot systems, to validate it in the real world. 
The AMAV is equipped with an Intel(R) T265 tracking camera for localization and an RGB camera with a FoV of 120 degrees for observation generation. 
Each BMAV is equipped with an IMU and downward-facing optical flow sensor.
Meanwhile, each BMAV mounts a $3cm \times 3cm$ AprilTag for recognition and observation generation \cite{wang2016apriltag} (Fig. \ref{setup}b). 
The AMAV adopts ArduPilot frameworks for motion control (Fig. \ref{setup}c). 
A motion capture system provides millimeter-level ground truth at 240 FPS in the experiment area of $8m \times 8m$ (Fig. \ref{setup}a and Fig. \ref{setup}d).
\name runs on a server featuring 128GB of memory and an Intel(R) Xeon(R) Gold 6242R CPU.
We test the robustness of TransformLoc on a physical-feature-based simulator in an experiment area similar to Fig \ref{setup}a. 

\begin{figure*}[t]
\setlength{\abovecaptionskip}{0.cm} 
\setlength{\belowcaptionskip}{-0.4cm} 
\setlength{\subfigcapskip}{-0.1cm}  
\centering
    \subfigure[The CDF of ATE]{
        \centering

            \includegraphics[width=0.63\columnwidth]{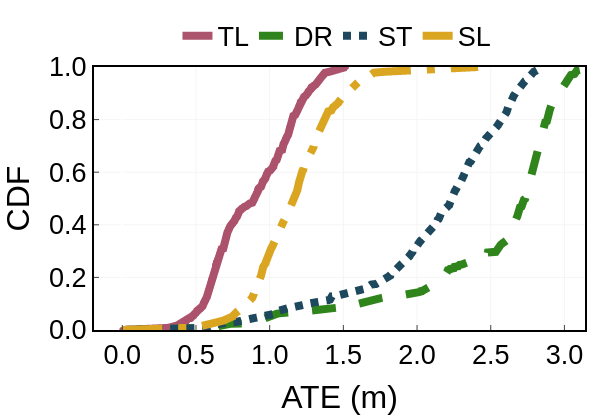}
    }
    \subfigure[Success rate within 200s]{
        \centering
            \includegraphics[width=0.63\columnwidth]{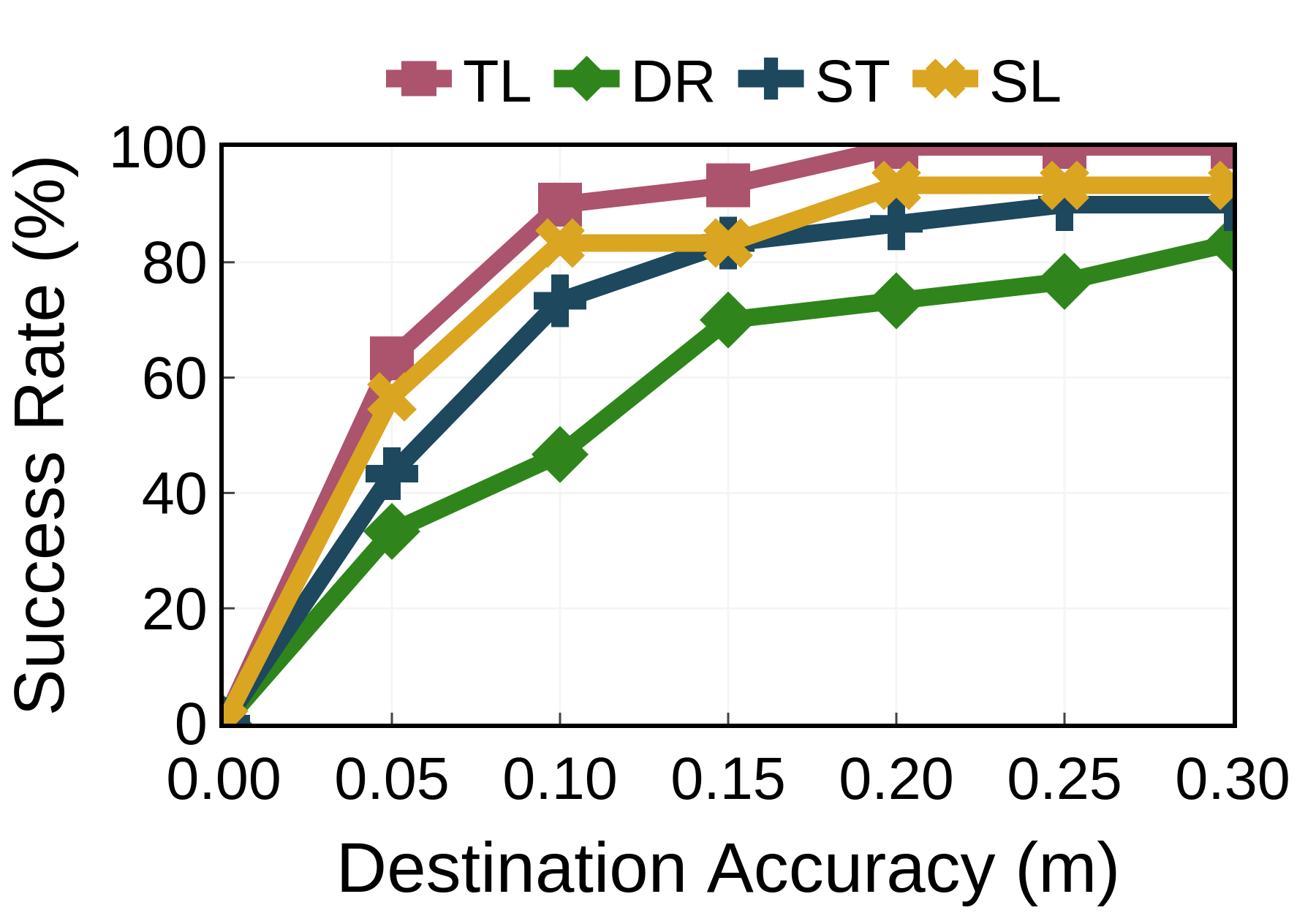}
    }
    \subfigure[Success rate with 0.2m accuracy]{
        \centering
            \includegraphics[width=0.63\columnwidth]{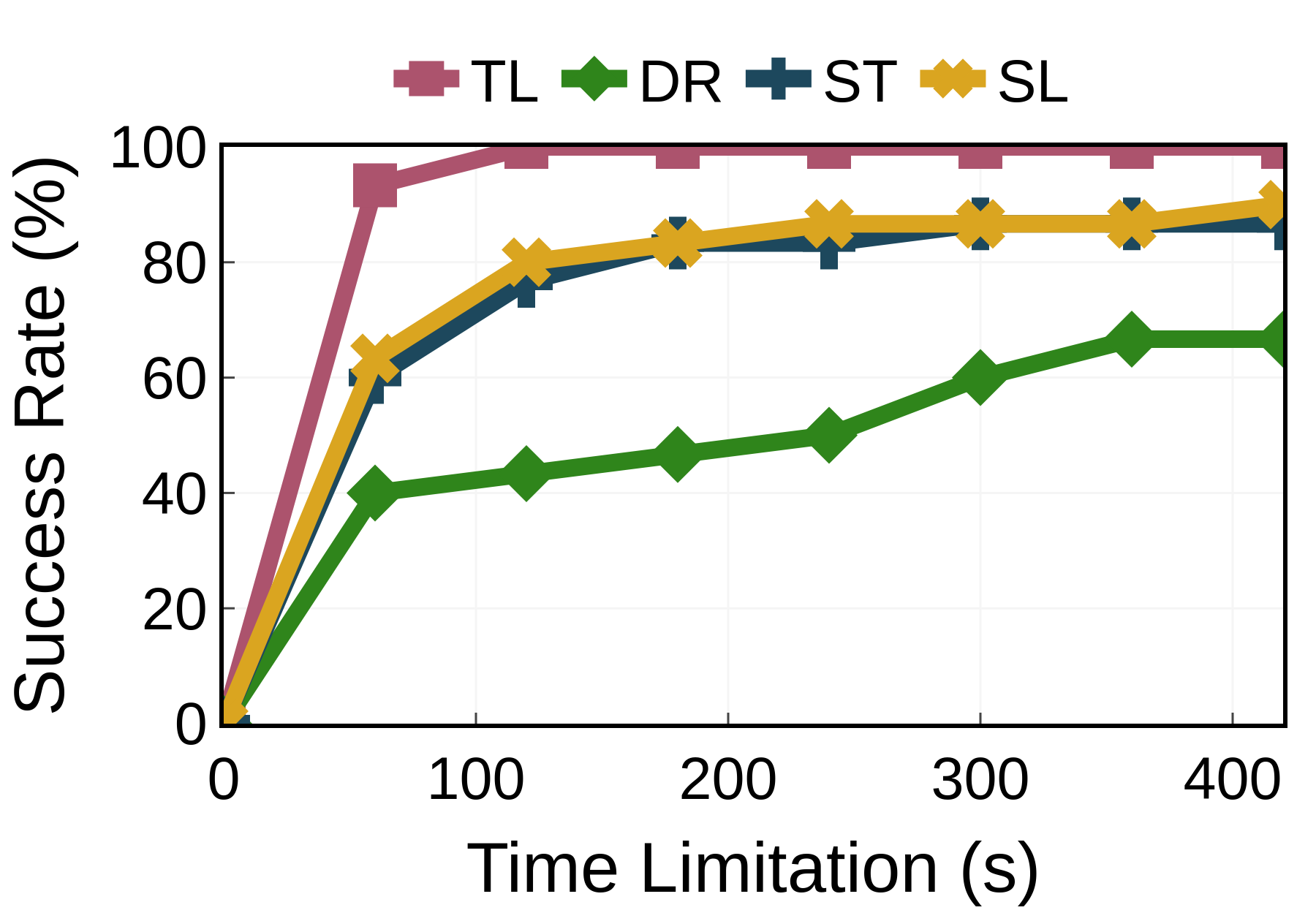}
    }
\caption{Overall performance in in-field experiments with two AMAVs and six BMAVs. Our TransformLoc outperforms the baselines, demonstrating superior performance in both localization error and navigation success rate.
}
\label{in-field}
\vspace{-0.4cm}
\end{figure*}

\begin{figure*}[t]
\setlength{\abovecaptionskip}{0.cm} 
\setlength{\belowcaptionskip}{-0.4cm} 
\setlength{\subfigcapskip}{-0.1cm}  
\centering
    \subfigure[The CDF of ATE]{
        \centering
            \includegraphics[width=0.63\columnwidth]{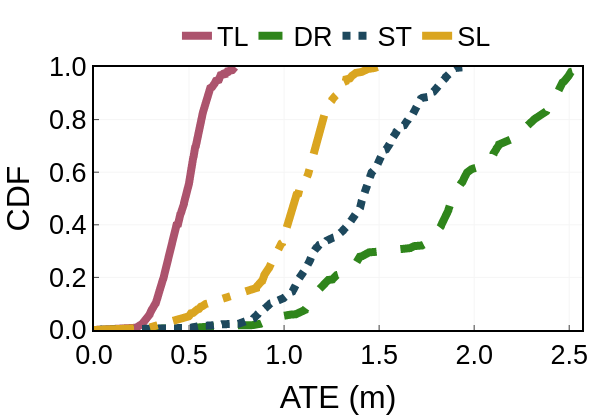}
    }
    \subfigure[Success rate within 200s]{
        \centering
            \includegraphics[width=0.63\columnwidth]{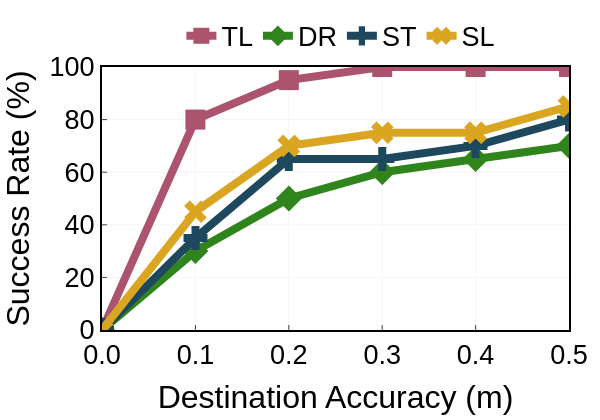}
    }
    \subfigure[Success rate with 0.2m accuracy]{
        \centering
            \includegraphics[width=0.63\columnwidth]{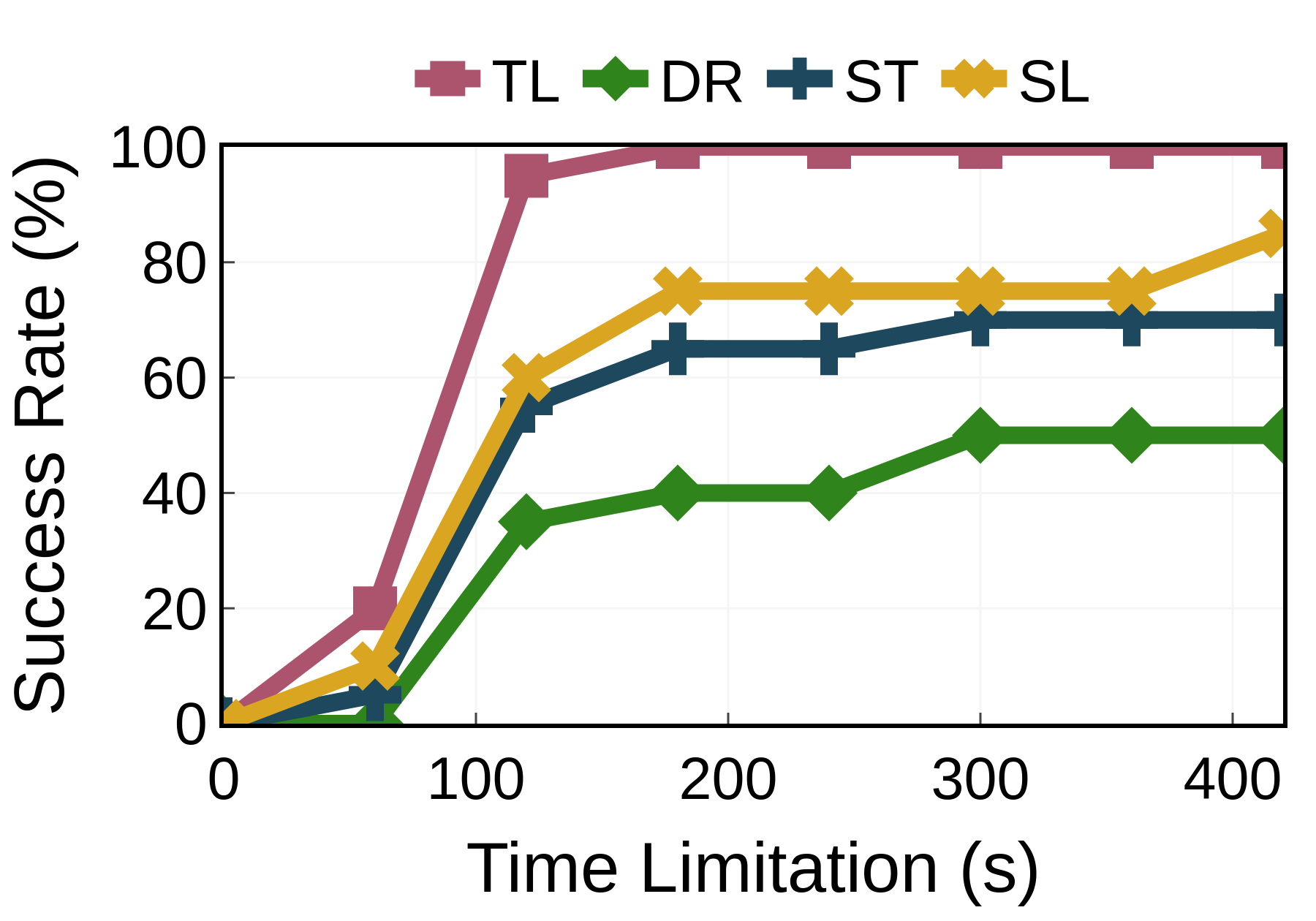}
    }
\caption{Overall performance in physical feature-based simulation with five AMAVs and twenty BMAVs. Even with the increased numbers of AMAVs and BMAVs, TransFormLoc continued to outperform the baseline in terms of localization accuracy and navigation success rate.
}
\label{simOverall}
\vspace{-0.4cm}
\end{figure*}

\begin{figure*}[t]
\setlength{\abovecaptionskip}{0.cm} 
\setlength{\belowcaptionskip}{-0.4cm} 
\setlength{\subfigcapskip}{-0.1cm}  
\centering
    \subfigure[Impact of the quantity of AMAVs]{
        \centering
        \includegraphics[width=0.62\columnwidth]{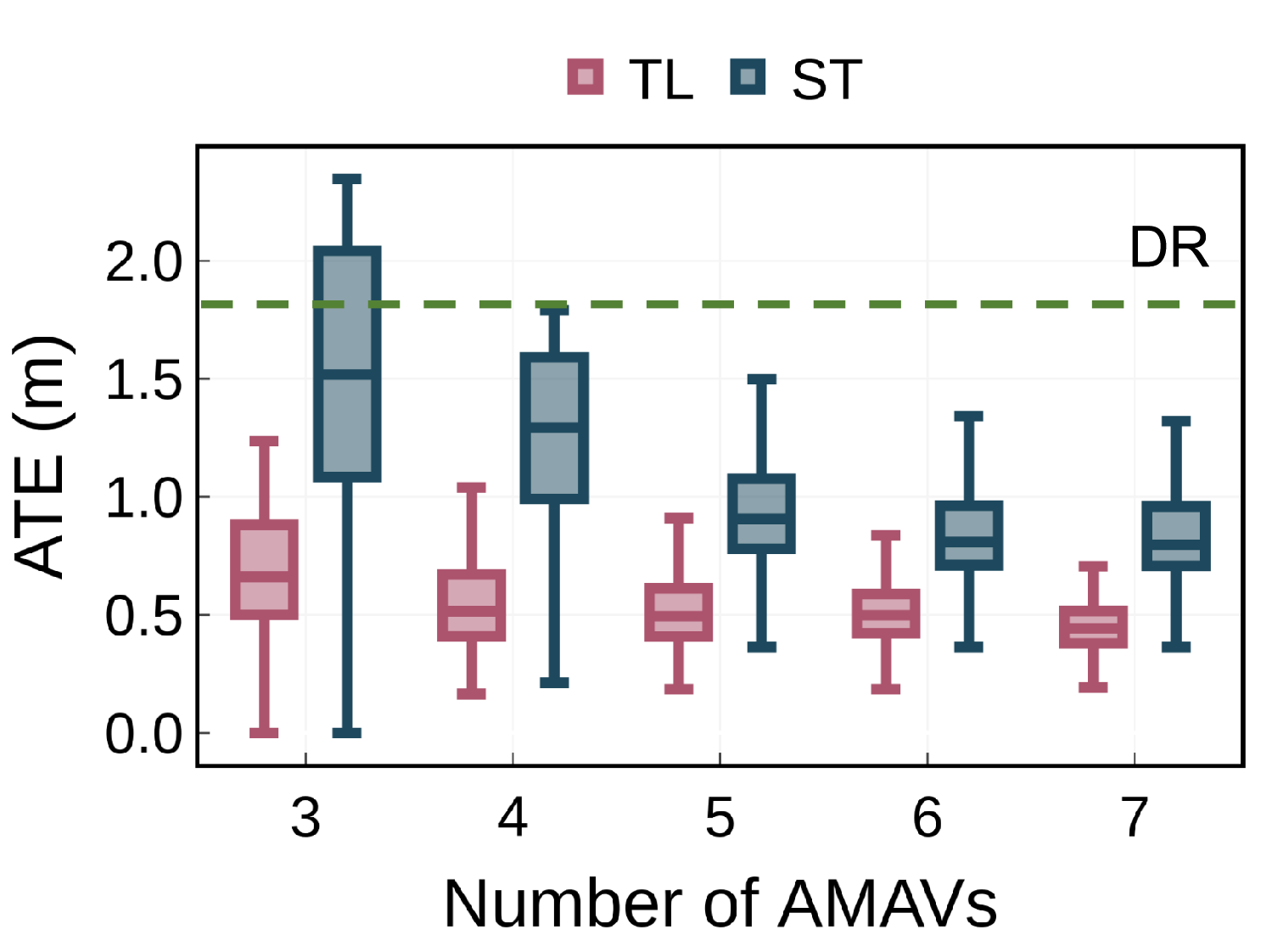}
    }
    \subfigure[Impact of the quantity of bMAVs]{
        \centering
            \includegraphics[width=0.67\columnwidth]{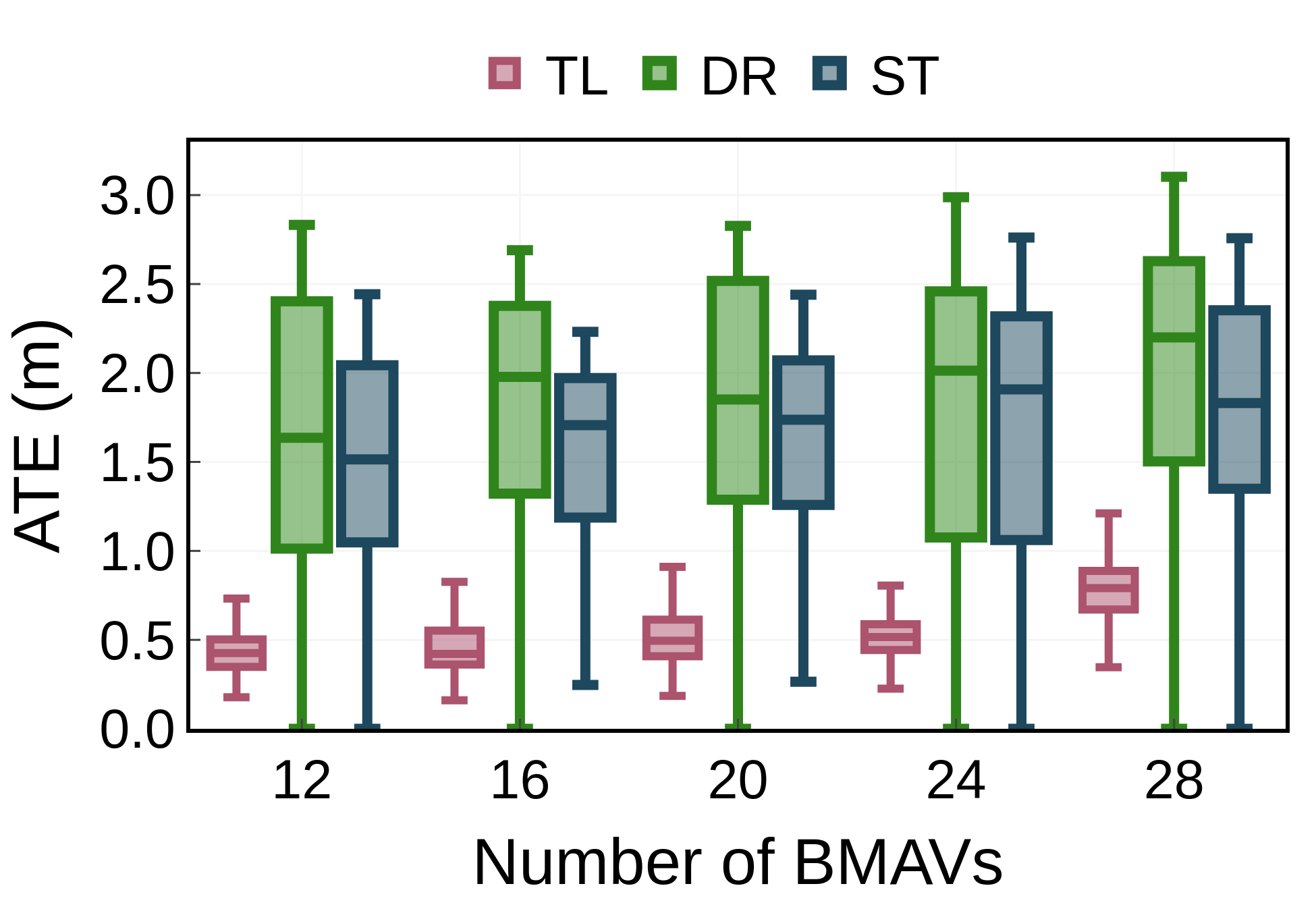}
    }
    \subfigure[Correlation of ATE and indicator]{
        \centering
            \includegraphics[width=0.6\columnwidth]{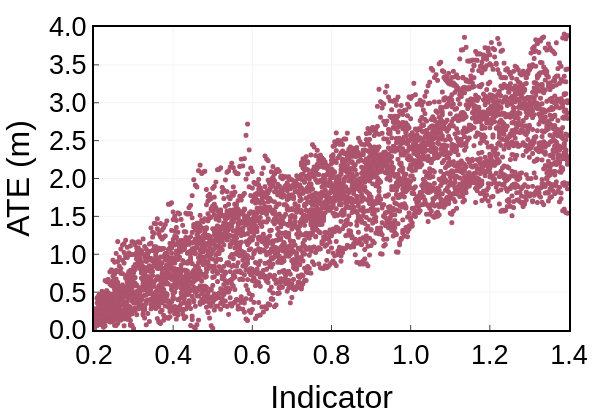}
    }
\caption{System Robustness Evaluation with five AMAVs and twenty BMAVs. The experimental results indicate that TransformLoc is capable of adapting to changes in the quantities of AMAVs and BMAVs. Moreover, our indicator can infer the localization errors of BMAVs.}
\label{simrobust}
\vspace{-0.4cm}
\end{figure*}

\textbf{Experiment setting.}
The BMAVs have a maximum velocity of 0.5$m/s$ and a command interval $\delta$ of 5 steps. 
The AMAVs have an observation angle $\theta$ of 120 degrees and maximum observation distance $r_m$ of 1\textit{m}. The control commands of AMAVs use motion primitives $\{(u, \omega)\}| u \in \{0, 1, 3\}m/s, \omega \in \{0, \pm1, \pm3\}rad/s$. 
We evaluated the performance by conducting in-field experiments with two AMAVs and six BMAVs, and simulations with five AMAVs and twenty BMAVs. 
The BMAVs move toward the edge of the room for environmental sensing. 
The simulations are conducted for $420$ seconds, consistent with the typical battery life of the DJI Robomaster TT utilized in in-field experiments. 
The standard deviation of noise models for BMAVs' motion, AMAVs' range and bearing measurements are determined based on in-field experiments and set to 20\%, 10\%, and 5\% of the measured values, respectively.

\textbf{Comparative Methods.} 
We tested TransformLoc (TL) against three baseline methods that do not require a localization infrastructure. 
\ding{172} Dead-Reckoning (DR) estimates the location of BMAVs with only measurements from motion sensors \cite{borenstein1996navigating}; 
\ding{173} Station (ST) utilizes AMAVs with fixed locations to generate observations for BMAVs \cite{xu2020learning};
\ding{174} H-SwarmLoc (SL) \cite{wang2022h} is a SOTA method that navigates one AMAV based on reinforcement learning to assist BMAVs in localization. 
To ensure a fair comparison, we standardize SL to group MAVs in a consistent manner and direct AMAV to generate observation for BMAVs within the same group during navigation \cite{wang2022h}.

\textbf{Evaluation Metrics.}
TransformLoc aims to improve the localization accuracy and navigation success rate of BMAVs. We use two metrics: 
\ding{172} Localization error: we compares the localization error of different BMAVs at each timestep, which is known as \textit{absolute trajectory error} (ATE); 
\ding{173} Success rate: this is measured by counting the ratio of BMAVs that reach destinations under constraints.

\subsection{Overall Performance} 
\textbf{In-field experiments.}
Fig. \ref{in-field} specifically focuses on in-field experimental results.
Regarding localization error, a 420-second random walk is conducted for BMAVs, and the cumulative distribution function (CDF) of the ATE is plotted.
TransformLoc achieves an ATE below $1.5m$, while the SL remains below $2.4m$, the ST below $2.8m$, and the DR below $3m$ (Fig. \ref{in-field}a).
Regarding navigation performance, the success rate of BMAVs over a 200-second duration increases for all methods as the destination accuracy decreases (Fig. \ref{in-field}b).
TransformLoc surpasses all baselines, achieving a remarkable 63\% success rate with a stringent destination accuracy constraint of 0.05m.
Furthermore, under looser destination accuracy limits (0.2m-0.3m), TransformLoc attains a 100\% navigation success rate, outperforming the baselines.
Additionally, as the time limitation increases, the success rate of BMAVs with a destination accuracy constraint of 0.2m improves for all methods, as they have more time to reach the target (Fig. \ref{in-field}c).
TransformLoc exhibits superior performance compared to the baselines, achieving a navigation success rate 22.3\% higher than SL, 25.6\% higher than ST, and 55.6\% higher than DR within strict time constraints (60-180 seconds).
The above results illustrate enhancements in performance achieved by TransformLoc as it efficiently allocates AMAVs' resources to bolster the localization and navigation capabilities of BMAVs.

\textbf{Physical feature-based simulation.}
Fig. \ref{simOverall} illustrates the performance of TransformLoc in the physical-feature-based simulation. 
In terms of localization error, TransformLoc achieves an impressive ATE below $0.7m$, while SL, ST, and DR maintain ATE below $1.5m$, $2m$, and $2.5m$ respectively (Fig. \ref{simOverall}a).
Regarding navigation success rate, TransformLoc consistently outperforms SL, ST, and DR by significant margins of 25\%, 26.67\%, and 33.3\% on average, respectively, with varying destination accuracy over a 200-second duration (Fig. \ref{simOverall}b).
Furthermore, under a destination accuracy of 0.2m, TransformLoc surpasses SL, ST, and DR by an average of 22.8\%, 25.6\%, and 42.5\% respectively, with varying time limitations (Fig. \ref{simOverall}c).
These results highlight the remarkable performance improvements achieved by TransformLoc.

\subsection{System Robustness Evaluation}

\textbf{Impact of Number of AMAVs.}
Fig. \ref{simrobust}a compares ATE of BMAVs against varying numbers of AMAVs with different methods. The TransformLoc exhibits lower errors than DR, which means the introduction of AMAVs reduced ATE. With the number of AMAVs increasing from 1 to 5, ATE of TransformLoc decreases below $0.8m$, significantly lower than ST which only helped BMAVs when passing over fixed AMAVs.
Overall, the results indicate that increasing the number of AMAVs significantly reduces ATE of BMAVs.

\textbf{Impact of Number of BMAVs.}
Fig. \ref{simrobust}b shows how different methods perform when localizing varying numbers of BMAVs. 
As the number of BMAVs increases from 12 to 28, the ATE of all methods increases due to fewer percentage of BMAVs passing through AMAVs. 
However, TransformLoc consistently achieves an ATE of less than $1.2 m$, which is more than 57\% less than any other baselines. 
These experimental results show that TransformLoc is able to allocate sensing resources of AMAVs to provide observations for different numbers of BMAVs, resulting in a low error. 

\textbf{Effectiveness of indicator.}
Fig. \ref{simrobust}c shows the effectiveness of our indicator.
The indicator ($tr(\Sigma_{j,t})$) grows with ATE of BMAV, indicating that this indicator reflects the extent of localization error.
\section{Related Work}\label{2}
\textbf{Multi-agent for environmental sensing.}
Advancements in AI have fueled extensive research into multi-agent systems, harnessing their parallel operational pipelines and coordinated complementarity to improve sensing coverage and reduce time requirements for sensing tasks \cite{wang2021lifesaving}. Agents in these systems can integrate various sensors, such as cameras \cite{xu2022swarmmap}, Radar \cite{lu2020milliego}, Lidar \cite{li2022motion}, IMU \cite{chen2015drunkwalk}, acoustic sensors \cite{wang2022micnest}, and gas sensors \cite{luo2023field}. This versatility enables the execution of diverse sensing tasks, including urban monitoring \cite{guo2021lightweight, chen2022deliversense}, hazardous gas sourcing \cite{liu2022fine}, and post-disaster data communication \cite{ren2023scheduling}. The data collected by these agents is shared through communication channels, providing insights into the agents' tasks and individual states, including motion policies \cite{chen2020adaptive, zhou2021intelligent}.

\textbf{Localization and navigation of MAV swarm.}
In the realm of sensing tasks, precise localization and navigation play a pivotal role in facilitating effective collaboration among MAV swarms, particularly given the inherent constraints in computing, communication, and sensing capabilities of individual MAVs \cite{li2022tract}. Simultaneous Localization and Mapping (SLAM) stands out as the most widely employed method \cite{mur2015orb}. This method entails outfitting MAVs with a suite of sensors, including RGB cameras, LiDAR, and depth cameras, to gather information about the environment and their own states. Subsequently, navigation algorithms are deployed to guide MAVs to their intended destinations \cite{xu2020edge, zhou2022swarm}.
While collaborative SLAM utilizing multiple MAVs has been investigated in prior studies \cite{xu2022swarmmap}, the prohibitive cost of sensors imposes limitations on the scalability of MAV swarms \cite{schmuck2019ccm}. In response to this challenge, approaches centered on external infrastructure have been proposed. For instance, radio frequency-based localization offers high accuracy at a relatively low cost of sensors \cite{chi2022wi}. However, these approaches rely on the presence of additional installed localization infrastructure, presenting challenges in environments where localization infrastructure may have been destroyed or where such installations are impractical, especially in hazardous conditions \cite{kumar2014accurate}.

To address these constraints, we propose a collaborative and adaptive localization framework, named \textit{\name}, for a heterogeneous MAV swarm. 
Within this framework, a group of AMAVs serves as mobile localization infrastructure, collaboratively sharing sensing and computing capabilities with a larger number of BMAVs. This cooperative approach enables precise localization for a MAV swarm at a more economical cost.

\section{Discussion}
We delve into several influential factors of TransformLoc.

\noindent $\bullet$ Communication Load: The BMAV transmits location estimates (mean and covariance) to AMAVs, and AMAV transmits motion commands to BMAVs. When catering to twenty BMAVs, an AMAV both receives and transmits less than 10KB of data every $\delta$ seconds. During grouping, an AMAV gathers location information from others and transmits grouping results to others. with twenty AMAVs, the data volume stays below 10KB. In summary, the communication load is deemed manageable.

\noindent $\bullet$ Localization accuracy of AMAV: 
The system improves the localization accuracy of BMAVs when the localization error of AMAVs is less than 10cm. Achieving this level of accuracy is easily feasible by outfitting an AMAV with a Lidar or depth camera and employing SLAM methods, such as VINS (Visual-Inertial Navigation System).
\section{Conclusion} \label{6}
In this paper, we propose the design of TransformLoc, a new framework that transforms AMAVs into mobile localization infrastructures. The innovation of TransformLoc lies in two aspects: 1) we derive an error-aware joint location estimation model to integrate inaccurate estimation from BMAVs with observations from AMAVs to perform joint location estimation of BMAVs, assisted by an uncertainty-aided inference method; and 2) we design a proximity-driven adaptive grouping-scheduling strategy to dynamically allocate AMAVs' sensing resources to assist BMAVs. The evaluation through in-field experiments and large-scale physical feature-based simulations demonstrate the superior performance of the TransformLoc.

\section{ACKNOWLEDGMENTS}
This paper was supported by the National Key R\&D program of China No. 2022YFC3300703, the Natural Science Foundation of China under Grant No. 62371269. Guangdong Innovative and Entrepreneurial Research Team Program No. 2021ZT09L197, Shenzhen 2022 Stabilization Support Program No. WDZC20220811103500001, and Tsinghua Shenzhen International Graduate School Cross-disciplinary Research and Innovation Fund Research Plan No. JC20220011.
We acknowledge the support from the Tsinghua Shenzhen International Graduate School-Shenzhen Pengrui Endowed Professorship Scheme of Shenzhen Pengrui Foundation.

\newpage
\appendices

\section{Key definitions}\label{A1}

\subsection{Environmental Description} 
Let $\Omega$ be a bounded space in $\mathbb{R}^3$ with length, width, and height $L$, $W$, and $H$, respectively, where the heterogeneous MAV swarm operates.
This swarm comprises a fixed number of BMAVs and AMAVs that operate independently. 
For simplicity, we assume that both types of MAVs operate at the same altitude in this paper.

\subsection{State model of AMAV} 
The heterogeneous MAV swarm contains $M$ ($M\textgreater1$) AMAVs. The state of each AMAV $A_j$ at time $t$ consists of the location and the motion command.
The location is denoted as $\boldsymbol{x_{j,t}}=(x_{j,t}^1, x_{j,t}^2, \phi_{j,t})$, where $\phi_{j,t}$ is the orient angle. The distance between two locations is $d_{\chi}$. For motion command, $\boldsymbol{u_{j, t}} = (u_{j,t}, \omega_{j,t})$, where $u_{j,t}$ and $\omega_{j,t}$ are the translational and rotational velocities, respectively.
The AMAVs estimate their location accurately with advanced sensing capabilities and follow the motion model $\boldsymbol{x_{j, t}} = f(\boldsymbol{x_{j, t-1}}, \boldsymbol{u_{j, t-1}})$, which can be represented as follows:
\begin{equation}
\left(\begin{array}{c}
x_{j,t}^1 \\
x_{j,t}^2 \\
\phi_{j,t}
\end{array}\right) =
\left(\begin{array}{c}
x_{j,t-1}^1 \\
x_{j,t-1}^2 \\
\phi_{j,t-1}
\end{array}\right) + \left(\begin{array}{c}
u_{j,t-1} \cos \left(\phi_{j,t-1}\right) \\
u_{j,t-1} \sin \left(\phi_{j,t-1}\right) \\
\omega_{j,t-1}
\end{array}\right).
\end{equation}

\subsection{State model of BMAV} 
The heterogeneous MAV swarm includes $N$ ($N\textgreater1$) BMAVs. The state of each BMAV $B_i$ at time $t$ consists of the location and the motion command.
The location of $B_i$ is $\boldsymbol{y_{i, t}}=(y_{i,t}^1, y_{i,t}^2)$, and its motion command is $\boldsymbol{v_{i, t}} = (v_{i,t}^1, v_{i,t}^2)$.
$B_i$ follows double integrator dynamics with Gaussian noise represented as follows: 
\begin{equation}
\begin{gathered}
\left(\begin{array}{l}
y_{i, t+\delta}^1 \\
y_{i, t+\delta}^2
\end{array}\right)=\left(\begin{array}{l}
y_{i, t}^1 \\
y_{i, t}^2
\end{array}\right)+\delta\left(\begin{array}{l}
v_{i, t}^1+n_{i, t}^1 \\
v_{i, t}^2+n_{i, t}^2
\end{array}\right),\\
\text{$n_{i, t}^1$ is drawn from $p(n^1)$,} \\
\text{$n_{i, t}^2$ is drawn from $p(n^2)$,}
\end{gathered}
\label{y}
\end{equation}
where $n_{i, t}^1$ and $n_{i, t}^2$ are drawn from motion noise models $p(n^1)$ and $p(n^2)$ respectively, $\delta$ ($\delta > 1$) denotes the time interval between two commands.
$p(n^1)$ and $p(n^2)$ empirically obtained from the testbed are specified as normal distributions with mean $\mu = 0$ and variance $\sigma$, expressed as a percentage of $v_{i, t}^1$ and $v_{i, t}^2$, same as outlined in \cite{chen2020h}.
The location estimation of $B_i$ is denoted as $\boldsymbol{\widehat{y}_{i, t}}$, and the covariance matrix is denoted as $\Sigma_{i, t}$.
The noisy motion measurement from $B_i$'s IMU is denoted as $\boldsymbol{\dot{v}_{i, t}}$.

\subsection{Observation model of AMAV}
The observation angle of the visual sensor is denoted by $\phi$, and the FoV of $A_j$ at time $t$ is defined as follows,
\begin{equation}
\begin{aligned}
& F_{j, t}=\{(x, y) \mid \\
& \quad \quad 0<\sqrt{\left(y-x_{j,t}^2\right)^2+\left(x-x_{j,t}^1\right)^2}<r_m \quad \bigcap \\
& \quad -\frac{\phi}{2}<\arctan\left(\left(y-x_{j,t}^2\right)\left(x-x_{j,t}^1\right)\right)-\phi<\frac{\phi}{2}\}, 
\end{aligned}
\label{F_t}
\end{equation}
where $r_m$ is the maximum distance that the sensor can observe, which is limited by the observation technique.

When $B_i$ in the FoV of $A_j$ at time $t$, $A_j$ generates noisy observation $z_{i, j, t}$ for $B_i$, which consists of range $r_{i, j, t}$ and bearing $\alpha_{i, j, t}$  for $B_i$, both $r_{i, j, t}$ and $\alpha_{i, j, t}$ are relative to $A_i$,
\begin{equation}
\begin{gathered}
\begin{aligned}
z_{i, j, t}
&=h(\boldsymbol{x_{j, t}}, \boldsymbol{y_{i, t}}) + \boldsymbol{n_{i, j, t}},\\
h(\boldsymbol{x_{j, t}}, \boldsymbol{y_{i, t}})
&:=\left[\begin{array}{c}
r_{i, j, t} \\
\alpha_{i, j, t}
\end{array}\right]\\
&:=\left[\begin{array}{c}
\sqrt{\left(y_{i, t}^1-x_{j, t}^1\right)^2+\left(y_{i, t}^2-x_{j, t}^2\right)^2} \\
\arctan \left(\left(y_{i, t}^2-x_{j, t}^2\right)\left(y_{i, t}^1-x_{j, t}^1\right)\right)-\phi
\end{array}\right]\\
\boldsymbol{n_{i, j, t}}
&:= [n_{i, j, t}^r, n_{i, j, t}^\alpha]^T
\end{aligned},\\
\text{$n_{i, j, t}^r$ is drawn from $p(n^r)$}, \\
\text{$n_{i, j, t}^\alpha$ is drawn from $p(n^\alpha)$}.
\end{gathered}
\label{z}
\end{equation}
The noise models for range and bearing measurements are denoted as $p(n^r)$ and $p(n^\alpha)$, respectively, and are empirically obtained from our testbed.
$p(n^r)$ and $p(n^\alpha)$ are specified as a normal distribution with mean $\mu = 0$ and variance $\sigma$, expressed as a percentage of $r_{i, j, t}$ and $\alpha_{i, j, t}$.

\subsection{Linearization of Observation model} 
We compute the Jacobian matrix of $h(\boldsymbol{x}, \boldsymbol{y})$ based on Taylor expansion by taking the gradient with respect to the location estimation of $\boldsymbol{y}$ to linearize the observation model,

\begin{equation}
\setlength\abovedisplayskip{3pt}
\setlength\belowdisplayskip{3pt}
\begin{aligned}
&\nabla_{\boldsymbol{y}} h(\boldsymbol{x}, \boldsymbol{y})\\
&=\frac{1}{r(\boldsymbol{x}, \boldsymbol{y})}\left[\begin{array}{cc}
\left(y^1-x^1\right) & \left(y^2-x^2\right) \\
-\sin \left(\phi+\alpha(\boldsymbol{x}, \boldsymbol{y})\right) & \cos \left(\phi+\alpha(\boldsymbol{x}, \boldsymbol{y})\right)
\end{array}\right].
\end{aligned}
\label{linearize}
\end{equation}

To simplify notation, we let
$\boldsymbol{y_t} := {[\boldsymbol{y_{1, t}^T}, \dots, \boldsymbol{y_{N, t}^T}]}^T$, 
$\boldsymbol{x_t} := {[\boldsymbol{x_{1, t}^T}, \dots, \boldsymbol{x_{M, t}^T}]}^T$, 
$\boldsymbol{u_t} := {[\boldsymbol{u_{1, t}^T}, \dots, \boldsymbol{u_{M, t}^T}]}^T$, 
$\boldsymbol{\widehat{y}_t} := {[\boldsymbol{\widehat{y}_{1, t}^T}, \dots, \boldsymbol{\widehat{y}_{N, t}^T}]}^T$, 
${\Sigma_{t}} := {diag(\Sigma_{1, t}^T, \dots, \Sigma_{N, t}^T)}^T$. 

\section{Problem Formulation}\label{A2}
We aim to optimize the trace of the covariance of BMAVs’ location estimation $\Sigma_t$ to limit the localization error within a finite time horizon $T$, considering the motion commands of each AMAV within $T$.
The mathematically formulated sensing resource allocation problem is given by,
\begin{equation}
\min _{\boldsymbol{u_0}, \ldots, \boldsymbol{u_{T-1}}} \xi(T) = \sum_{t=1}^T\left(tr\left(\Sigma_t\right)\right),
\end{equation}
\text {\centerline{s.t.} }
\begin{align}
    & \boldsymbol{x_{t+1}} = f(\boldsymbol{x_{t}}, \boldsymbol{u_{t}}), t \in \{0,\dots, T-1\} \\
    & 0 \leq x_{j, t}^1, y_{i, t}^1 \leq L, \\
    & \quad j \in \{0,\dots, N\}, i \in \{0,\dots, M\}, t\in\{0,\dots, T-1\} \notag \\
    & 0 \leq x_{j, t}^2, y_{i, t}^2 \leq W, \\
    & \quad j \in \{0,\dots, N\}, i \in \{0,\dots, M\}, t\in\{0,\dots, T-1\} \notag \\
    & \Sigma_{t+1}=\rho_{t+1}^e\left(\rho_t^p\left(\Sigma_t\right), \boldsymbol{x}_{t+1}\right), t\in\{0,\dots, T-1\}.
\label{problem}
\end{align}
$\rho_t^e$ and $\rho_{t+1}^p$ are correction and prediction steps, which are described in Section III-B.

\newpage
\bibliographystyle{IEEEtran}
\bibliography{infocom}

\vfill

\end{document}